\documentclass[authoryear,review]{elsarticle}
\usepackage{graphicx}%
\usepackage{multirow}%
\usepackage{amsmath,amssymb,amsfonts}%
\usepackage{amsthm}%
\usepackage{mathrsfs}%
\usepackage[title]{appendix}%
\usepackage{xcolor}%
\usepackage{textcomp}%
\usepackage{manyfoot}%
\usepackage{booktabs}%
\usepackage{algorithm}%
\usepackage{algorithmicx}%
\usepackage{algpseudocode}%
\usepackage{listings}%
\usepackage{verbatim}
\usepackage{enumitem}
 \usepackage{comment}
 \usepackage{epigraph}

\begin{document}

\begin{frontmatter}

\title{When correcting for regression to the mean is worse than no correction at all}

\author[mymainaddress]{Jos\'e F.  Fontanari}
\ead{fontanari@ifsc.usp.br}

\address[mymainaddress]{Instituto de F\'{\i}sica de S\~ao Carlos,  Universidade de S\~ao Paulo,  13566-590 S\~ao Carlos,  S\~ao Paulo,  Brazil}

\author[Spain,Portugal]{Mauro Santos}%\corref{mycorrespondingauthor}}
\ead{mauro.santos@uab.es}

\address[Spain]{Departament de Gen\`etica i de Microbiologia, Grup de Gen\`omica, Bioinform\`atica i
Biologia Evolutiva (GBBE), Universitat Aut\`onoma de Barcelona, Spain}
\address[Portugal]{cE3c - Centre for Ecology, Evolution and Environmental Changes \& CHANGE - Global
Change and Sustainability Institute, Lisboa, Portugal}

\begin{abstract}
The ubiquitous regression to the mean (RTM) effect complicates statistical inference regarding the relationship between baseline levels of a biological variable and its subsequent change. We demonstrate that common RTM correction methods are problematic: the  Berry et al. method,  popularized  by Kelly \& Price in The American Naturalist,  is unreliable for hypothesis testing or effect-size estimation, leading to systematic bias and inflated error rates. Conversely, while the Blomqvist method is theoretically unbiased, its high sampling variance limits its practical utility in small-to-moderate datasets.  Using a structural linear model, we show that the most robust approach to navigating RTM is not to  correct  the data, but to evaluate the uncorrected crude slope against a structural null expectation derived from measurement repeatability---the proportion of total variance attributable to true individual differences.  We illustrate this approach using empirical data from studies on lizard thermal physiology and bird telomere dynamics. Ultimately, we argue that any conclusion regarding a differential treatment effect is statistically unfounded without a clear understanding of the experiment’s repeatability. 
\end{abstract}

\end{frontmatter}

\newpage{}

\section*{Introduction}

In ecology and physiology, a frequent objective is to understand how an individual's initial state influences its subsequent response to an environmental stimulus or treatment. For example, researchers may ask if lizards with low basal heat tolerance show a greater increase in tolerance after acclimation compared to those already near their physiological limit (\citealt{Deery_2021}). Similarly, an ecologist might investigate if a bird’s initial body mass predicts the magnitude of its mass loss during incubation (\citealt{Cichon_1999}).

This type of question is typically addressed by analyzing the change score---the difference ($d = x_2 - x_1$) between a post-test ($x_2$) and a pre-test ($x_1$) measurement---as a function of the initial value. While some statistical disciplines prefer modeling the post-test value with the pre-test as a covariate (ANCOVA) to assess treatment effects (\citealt{Harrell_2015}), the choice between these two approaches has been a source of profound statistical tension for decades. This tension is famously exemplified by Lord’s Paradox (\citealt{Lord_1967}), which demonstrates how change scores and ANCOVA can provide contradictory conclusions in observational data. In organismal ecology, however, the change score remains the standard framework because it directly maps onto biological concepts such as compensatory growth or  catch-up responses (\citealt{Metcalfe_2001}).

While common, assessing the relationship between change and initial values through correlation or regression is notoriously difficult due to two overlapping phenomena: mathematical coupling and regression to the mean. 
Mathematical coupling (\citealt{Archie_1981}) occurs because the independent variable ($x_1$) is arithmetically a component of the dependent variable ($d = x_2 - x_1$). This creates a structural dependence where even perfectly measured, independent variables will exhibit a negative correlation (\citealt{Pearson_1897}).  When $x_1$ also contains measurement error, this coupling further manifests as regression to the mean (RTM), creating a  negative association that does not reflect any underlying biological process.  While mathematical coupling can be addressed with randomization tests (\citealt{Jackson_1991}),  these tests specifically evaluate a null hypothesis of complete independence between $x_1$ and $x_2$.  Our focus here is on the problem of RTM  (\citealt{Galton_1886}). This phenomenon,  also known as the  law of initial values  in physiological and psychological studies of response to a stimulus (\citealt{Wilder_1967, Geenen_1993}),  arises because when individuals are selected based on an extreme initial value, their subsequent measurement is  likely to be closer to the population mean simply because the extreme error in the first measurement is unlikely to be repeated  (\citealt{Galton_1886}).

Without a rigorous framework to distinguish these artifacts from true biological signals, researchers risk misinterpreting statistical noise as evidence of a differential treatment effect. We stress that the challenge of baseline measurement error is universal, though its manifestation depends on the chosen framework. In change-score analyses, this error drives both mathematical coupling and RTM. In the ANCOVA framework, while mathematical coupling is avoided, the RTM effect manifests as regression dilution (or attenuation), where measurement error in the baseline covariate leads to an underestimation of the relationship between variables and potentially biased treatment effects.

A number of alternative statistical methodologies have been proposed to address the challenges of assessing the relationship between change and initial values through correlation or regression,  particularly in the psychological (\citealt{Nesselroade_1980}) and clinical (\citealt{Chiolero_2013}) literature.  Recently, latent change score modeling, a type of structural equation modeling, has been extensively applied in psychological research (\citealt{Ferrer_2010}).  However, when only two data sets are available (e.g., pre- and post-test), this approach is also susceptible to RTM (\citealt{Sorjonen_2023}).  In biological research, the work of \citet{Berry_1984}, and its subsequent application and popularization by \citet{Kelly_2005}  in The American Naturalist (a paper cited over 177 times to date), has been particularly influential in highlighting the RTM problem.

The potential for misinterpretation is well-documented. As \citet{Forstmeier_2017}  cautioned: ``There is one final statistical phenomenon that we would like to highlight: `regression to the mean’\dots it is a sufficiently common trap and has led to errors in a wide range of scientific disciplines\dots Moreover, since the regression to the mean will consistently produce a spurious but often significant effect, and since we typically publish when encountering something significant, one can readily find erroneous interpretations of this artefact in the literature." \citet{Mazalla_2022}  and \citet{Slessarev_2023} provide recent examples of these statistical pitfalls in ecology. Consequently, there is a perceived imperative to correct for RTM. However, as our analysis will demonstrate, the adjustment method proposed by \citet{Berry_1984} and popularized by \citet{Kelly_2005} possesses structural limitations that lead to biased estimates under many biological scenarios.  This estimator appears to have been adopted in the ecological literature despite a lack of formal evaluation regarding its performance across different ratios of measurement error to between-subject variance ($\gamma^2$). Consequently, its application may lead to unreliable biological conclusions. 

From our experience, navigating the vast literature on the relationship between an initial value and a subsequent change of a continuous variable can be frustrating due to inconsistent terminology and a lack of a unified framework. To address this, we provide a reformulation and extension of key seminal articles, particularly that by \citet{Hayes_1988}, to provide a clear path forward. Our analysis focuses specifically on the estimation of the regression slope of change on initial value, based on the understanding that the RTM effect is an inherent consequence of measurement error.

In this paper, we provide a rigorous evaluation of the statistical methods used to account for RTM. We first develop a formal linear structural model that explicitly incorporates measurement error and stochastic variation in treatment effects. Using this framework, we analytically evaluate the performance of the correction proposed by \citet{Berry_1984}. We demonstrate that this method remains sensitive to the underlying distribution of measurement error, which can lead to systematic bias and an increased risk of both Type I and Type II errors. We then compare these results with the estimator proposed by \citet{Blomqvist_1977}. Using computer simulations, we show that while the Blomqvist slope is theoretically unbiased, its high sampling variance in small-to-moderate datasets often limits its practical utility compared to the uncorrected `crude' slope. By identifying these trade-offs between bias and precision, we provide a clearer foundation for choosing an appropriate inferential strategy.

Finally, we argue that the most robust inferential strategy is not to 'correct' the observed data—a process that often introduces new biases—but to evaluate the uncorrected 'crude' slope within a structural null-testing framework. We provide a bootstrap-based method that allows researchers to determine if their observed results are statistically inconsistent with a null hypothesis of no differential treatment, even when measurement repeatability is not precisely known. We illustrate the practical utility of this approach using empirical datasets from studies on lizard thermal physiology and bird telomere dynamics. By providing a mathematically grounded guide for analyzing the relationship between change and initial values, we offer a path forward that preserves the biological intuition of the change score while maintaining the statistical rigor required to navigate the pitfalls of regression to the mean.

\section*{A Framework for Change}

\subsection*{A Structural Model of Change}

Consider a population of individuals (e.g., animals, plants, or experimental units) where we measure a continuous trait at two points in time. Let $X_1$ and $X_2$ represent the true (but unobservable) values of the trait before and after a treatment or an interval of time. To understand the regression to the mean (RTM) effect, it is essential to define a formal structural model that relates $X_2$ to $X_1$.

We assume that the true initial values $X_1$ in the population are normally distributed $X_1 \sim N(\mu, \gamma^2)$, where $\mu$ is the population mean and $\gamma^2$ represents the between-subject variance. Following \citet{Hayes_1988}, we model the true change $D = X_2 - X_1$ as a linear function of the initial value:
\begin{equation}\label{X2}
D = \alpha + \beta X_1 + \zeta,
\end{equation}
where the term $\alpha + \beta X_1$ represents the deterministic treatment effect. Here, $\beta$ is the parameter of primary interest, representing the \textit{differential treatment effect}---the degree to which the treatment's impact depends on the individual's initial state. The term $\zeta$ represents a stochastic effect, modeled as noise, with $\zeta \sim N(0, \nu^2)$, where $\nu^2$ quantifies the between-subject variation in the treatment's impact.

If a treatment affects all individuals additively and equally, then $\beta = 0$, and any differential effect is due to the stochastic noise $\zeta$. Conversely, a non-zero $\beta$ indicates a systematic relationship where the treatment varies predictably with the initial value. This structural framework allows us to distinguish between deterministic biological signals ($\beta$) and the statistical artifacts of RTM.

The true values $X_1$ and $X_2$ are not directly observable. Instead, we measure $x_1$ and $x_2$, which are subject to within-subject variation $\epsilon$ (encompassing both measurement error and short-term biological fluctuations)
\begin{eqnarray}
x_1 & = & X_1 + \epsilon_1  \label{xx1} \\
x_2 & = & X_2 + \epsilon_2,   \label{xx2}
\end{eqnarray}
where $\epsilon_i \sim N(0,\delta^2)$ and is assumed to be independent of the true values ($X$) and the biological noise ($\zeta$).  Here, $\delta^2$ represents the within-subject (error) variance, which accounts for both technical measurement error and the transient biological `noise' that occurs at the moment of sampling.  From these definitions, we derive the statistical properties of the measured values:
\begin{eqnarray}
\mathbb{E}(x_1) = \mu; & \quad & \mathbb{V}(x_1) = \gamma^2 + \delta^2 \\
\mathbb{E}(x_2) = (1+\beta)\mu + \alpha; & \quad & \mathbb{V}(x_2) = (1+\beta)^2 \gamma^2 + \nu^2 + \delta^2 \\
\mbox{cov}(x_1,x_2) & = & (1+\beta) \gamma^2
\end{eqnarray}
where $\mu$ is the population mean of the initial state, $\gamma^2$ is the between-subject variance of the initial true state (the structural variance), and $\nu^2$ represents the variance of the biological noise $\zeta$---the stochastic component of actual phenotypic change independent of the initial state.   These derived properties are fundamental for evaluating methods designed to correct for RTM. Consistent with standard frameworks for measurement error and RTM (\citealt{Galton_1886,Kelly_2005}), we assume that the latent variables and error terms are normally distributed. While the second-order moments defined here are general, the specific linear manifestations of RTM and the efficacy of the estimators discussed hereafter are most formally derived within this Gaussian framework.

A key parameter emerging from this structural model is measurement repeatability ($R$), defined as the proportion of the total observed variance that is attributable to true differences between individuals:
\begin{equation}\label{R_def}
R = \frac{\gamma^2}{\gamma^2 + \delta^2} = \frac{\mathbb{V}(X_1)}{\mathbb{V}(x_1)}.
\end{equation}

In biological terms, $R$ represents the reliability of a single measurement. When the measurement error is zero ($\delta^2 = 0$), $R = 1$ and the observed $x_1$ is identical to the true state $X_1$.  As the measurement error $\delta^2$ grows relative to the between-subject variance $\gamma^2$, the repeatability $R$ decreases, thereby obscuring the true biological signal.  This metric is central to the problem of RTM because the expected crude regression slope between observed change and initial value, under a null structural effect ($\beta = 0$), is $R - 1$ (\citealt{Hayes_1988}). Consequently, any observed negative correlation cannot be interpreted as a biological signal unless it is shown to be significantly more negative than the bias dictated by the measurement's own repeatability.

%----------------------------------------------------------
\begin{figure}[h!]
\centering
    \includegraphics[width=\textwidth]{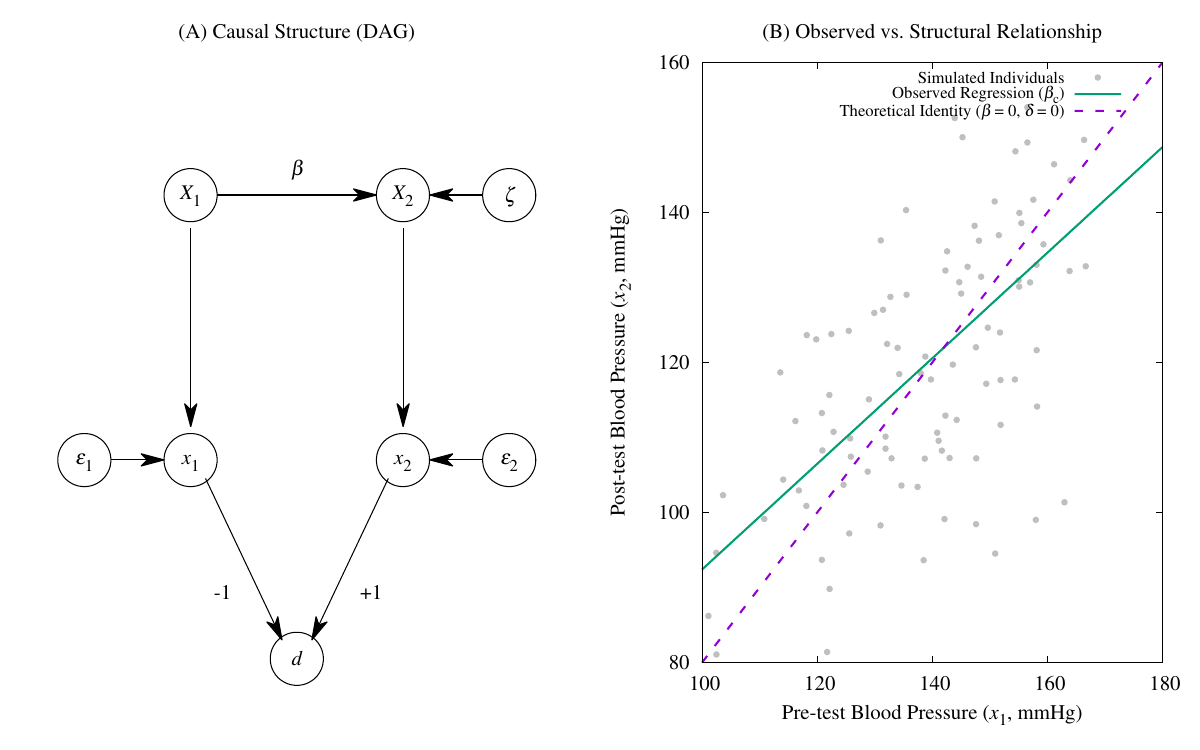}
\caption{ Conceptual framework of the structural model and Regression to the Mean (RTM). (A) A Directed Acyclic Graph (DAG) illustrating the causal paths between true states ($X$) and measured values ($x$). The structural estimand $\beta$ represents the true biological effect of the initial state on change.  Measurement errors ($\epsilon_1, \epsilon_2$) and biological noise ($\zeta$) create the observed values, while the change score $d$ is derived mathematically. (B) A simulation of $N=100$ individuals where the true structural effect is zero ($\beta = 0$). Parameters are based on a systolic blood pressure model system: $\mu = 141$ mmHg, between-subject SD $\gamma = 13.6$ mmHg, within-subject SD $\delta = 9.1$ mmHg, and stochastic biological noise $\nu = 10$ mmHg. These values result in a repeatability of $R \approx 0.69$.   The dashed purple line represents the theoretical identity line (unit slope), corresponding to the idealized case where both the systematic effect and measurement error are null ($\beta = 0, \delta = 0$). The solid green line represents the observed regression ($\beta_c$), which is tilted due to RTM. This illustrates how measurement error alone creates a misleading statistical association even in the absence of a systematic biological trade-off.
}
\label{fig:1rev}
\end{figure}
%----------------------------------------------------------

To provide a causal framework for our analysis, we use a Directed Acyclic Graph (DAG) in Figure \ref{fig:1rev}A. In a DAG, variables (nodes) are connected by arrows that represent hypothesized causal directions. These diagrams are particularly useful for highlighting how mathematical coupling creates spurious associations. Our DAG identifies the structural estimand, $\beta$, and illustrates how measurement error ($\epsilon_1$) enters the system. Because the change score $d$ is a derived variable ($d = x_2 - x_1$), it is mathematically coupled with $x_1$ via the shared error term $\epsilon_1$. This shared component creates a non-causal statistical association that biases the observed relationship between $x_1$ and $d$. As shown in Equation (\ref{R_def}), $\epsilon_1$ is the mechanical source of low repeatability ($R$); as the variance of this error increases relative to the true individual variance, this coupling effect grows stronger and $R$ decreases.   Note that while the intercept $\alpha$ shifts the mean of $X_2$, it is omitted from the DAG as it is a constant term and does not contribute to the covariance structure.  
In Figure \ref{fig:1rev}B, we illustrate the mechanics of RTM using parameters for systolic blood pressure derived from \citet{Gardner_1973} and \citet{Hayes_1988}. While clinical in origin, these parameters serve as a high-precision model system to illustrate the universal statistical properties of measurement error that affect any continuously distributed trait in ecology, such as body mass or physiological tolerance.
By setting $\beta = 0$, we simulate a scenario with a null structural effect. To provide a visual benchmark, the dashed purple line represents the theoretical identity line (a slope of 1), which depicts the expected relationship if both the structural effect and measurement error were null ($\beta = 0, \delta = 0$). Any deviation of the observed regression line (solid green line) from this unit slope is purely an artifact of measurement error ($\epsilon$), which manifests as the RTM effect. This visualization demonstrates how a null biological effect can still result in a statistical association that mimics a treatment effect due to the presence of measurement noise in the observed data.

\subsection*{The Bivariate Normal Framework}

It is instructive to compare this structural model with the common alternative, which assumes  measured values $x_1$ and $x_2$ are drawn from a bivariate normal (BVN) distribution (\citealt{Berry_1984, Kelly_2005})
\begin{equation}\label{Berry}
\left(\begin{array}{c}x_1\\x_2 \end{array}\right) \sim N \left[\left(\begin{array}{c}\mu_1 \\ \mu_2\end{array}\right), \left(\begin{array}{cc}\sigma^2_1 & \rho \sigma_1\sigma_2 \\ \rho\sigma_1\sigma_2 & \sigma^2_2 \end{array}\right)\right] 
\end{equation}
where $\mu_1$ and $\mu_2$ are the observed means, $\sigma^2_1$ and $\sigma^2_2$ are the observed phenotypic variances at times 1 and 2, and $\rho$ is the Pearson correlation coefficient between the measurements. These phenotypic parameters are related to the underlying structural model as follows: $\mu_1 = \mu$, $\mu_2 = \mu + \alpha + \beta \mu$, $\sigma_1^2 = \gamma^2 + \delta^2$, $\sigma_2^2= (1+\beta)^2 \gamma^2 + \nu^2 + \delta^2$, and $\rho\sigma_1\sigma_2 = (1+\beta)\gamma^2$.

While mathematically equivalent to the BVN framework, our structural model provides a more transparent mapping between biological processes and observed data. This clarity is essential for identifying where standard statistical interpretations fail. In the following subsections, we analyze four critical areas where the reliance on observed correlations and variances leads to structural misinterpretations: (1) the discrepancy between structural parameters and observed estimates, (2) the limitations of the equal variance assumption, (3) the confounding effect of mathematical coupling, and (4) the systematic bias present in the crude regression slope.

\subsection*{Structural Parameters vs.  Observed Estimates}

In the bivariate normal framework, researchers often assume that a non-differential treatment effect---where all individuals respond identically regardless of their starting state---corresponds to a null correlation ($\rho = 0$) between pre- and post-test measurements ($x_1$ and $x_2$). This assumption, though common in some literature (e.g., \citet{Jackson_1991,Cichon_1999,Deery_2021}), is inconsistent with the structural model of change.  As shown in the DAG (Figure \ref{fig:1rev}A), a null structural effect is defined by $\beta = 0$. Under this condition, the expected relationship between the latent variables $X_1$ and $X_2$ is a parallel shift with a slope of 1. This idealized relationship---representing the scenario where both the biological effect and measurement error are absent---is depicted by the theoretical identity line (the dashed line in Figure \ref{fig:1rev}B).  However, the observed correlation coefficient under this structural null is not zero, but rather
\begin{equation}
\rho^* = \frac{\gamma^2}{\sqrt{(\gamma^2 + \delta^2)(\gamma^2 + \delta^2 + \nu^2)} }.
\end{equation}
This demonstrates that even when the biological effect is uniform ($\beta=0$), measurement error ($\delta$) and biological noise ($\nu$) induce a non-zero observed correlation. Consequently, testing against the conventional null hypothesis of $\rho=0$ is likely to lead to a Type I error, where the researcher identifies a  significant differential effect that is actually a statistical artifact of RTM.  A major practical challenge is that this structurally correct null hypothesis ($\rho^*$) depends on the measurement error variance $\delta^2$. While the total observed variances $\mathbb{V}(x_1)$ and $\mathbb{V}(x_2)$ are easily calculated, $\delta^2$ cannot be estimated from just two time points. This creates a methodological impasse: the easily testable hypothesis ($\rho=0$) is structurally incorrect, while the structurally correct hypothesis is not directly testable without prior knowledge of the measurement repeatability. This situation mirrors the RTM correction proposed by \citet{Blomqvist_1977}, which similarly requires an external estimate of $\delta^2$ to be effective (\citealt{Chiolero_2013}).
This impasse highlights that the conventional null hypothesis ($\rho=0$) is a descriptive property of the sample, whereas our structural model identifies the causal parameters required to understand the treatment effect.

\subsection*{Evaluation of the Equal Variance Assumption}

A frequent approach in the literature involves testing the equality of pre- and post-treatment variances ($\mathbb{V}(x_1)= \mathbb{V}(x_2)$) to identify a differential treatment effect.   This methodology originates with the concepts described by \citet{Galton_1886}  and frequently  employs Pitman’s test \citep{Pitman_1939}  to evaluate whether the variance ratio deviates from unity.\citep{Berry_1984,Chiolero_2013,Kelly_2005}. 
However, our structural model suggests that the equality of variances is an insufficient criterion for identifying the presence of $\beta$.  As shown in the DAG (Figure \ref{fig:1rev}A), the post-treatment variance is influenced not only by the starting state but also by stochastic biological noise ($\zeta$). Even under a null structural effect ($\beta = 0$), the variance ratio is
\begin{equation}
\frac{\mathbb{V}(x_2)}{\mathbb{V}(x_1)} = 1 + \frac{\nu^2}{\gamma^2 + \delta^2} = 1 + \frac{\nu^2}{\mathbb{V}(x_1)}.\end{equation}
Because biological noise contributes additional variance ($\nu^2 > 0$), the post-treatment variance will naturally exceed the pre-treatment variance unless a negative $\beta$ counteracts this increase.  Consequently, the assumption that $\mathbb{V}(x_1) = \mathbb{V}(x_2)$ represents a null biological state is structurally inconsistent with a model that accounts for individual-specific variation ($\zeta$).  Much like the correlation coefficient discussed in the previous section, the variance ratio depends on an unobservable parameter---in this case, the variance of the biological noise ($\nu^2$). Without an independent estimate of this noise from more than two time points, a variance ratio of 1 cannot be used as a reliable baseline for testing $\beta$.   While a ratio near unity is consistent with a null structural effect in a system with negligible biological noise, it could equally result from a negative $\beta$ that happens to offset the variance added by $\nu^2$ . Therefore, concluding that a treatment effect is non-differential simply because variances remain stable is as structurally unsupported as assuming that $\rho=0$ implies $\beta=0$.

\subsection*{Mathematical Coupling and the Change-Score Correlation}

In our framework, we define the true structural change as $D = X_2 - X_1$, and the slope $\beta$ captures its dependence on the initial state.  A negative value for $\beta$ indicates that subjects with higher initial values experience a smaller increase or a greater reduction in the trait compared to those with lower initial values (e.g., individuals with higher initial blood pressure experiencing a more pronounced drop, or those with higher initial heat tolerance showing a smaller temporal increase in thermal tolerance following sub-lethal exposure, known as the hardening response).
 
 The central challenge is that the observed slope, $\beta_c$ (from the regression of $d$ on $x_1$), systematically deviates from the structural parameter $\beta$ due to two distinct but related phenomena:
 \begin{enumerate}
 \item Mathematical Coupling: This is the  common variable  problem first identified by \citet{Pearson_1897}. Because $x_1$ is used both as the independent variable and as a component of the dependent variable ($d = x_2 - x_1$), a correlation is arithmetically  forced into the measurement. While previous efforts have focused on removing this spurious correlation through alternative null hypotheses \citep{Archie_1981, Kronmal_1993, Santos_2025}, these corrections do not resolve the underlying structural bias.
 \item The RTM Effect: Unlike coupling, which is a matter of arithmetic, RTM is a consequence of measurement error ($\delta$). As demonstrated in Figure \ref{fig:1rev}B, even if one were to account for the mathematical coupling problem, the presence of $\delta$ ensures that the observed slope $\beta_c$ will always be shallower than the true biological effect (the structural slope $\beta$).  Distinguishing between these two effects is vital. Mathematical coupling is a property of how we calculate change, whereas RTM is a property of how we measure the world. By framing the analysis within the structural model, it becomes clear that neither a zero correlation ($\rho=0$) nor a simple correction for coupling can recover the true value of $\beta$ without accounting for the measurement error variance.
 \end{enumerate}
 
The perceived mathematical overlap between these two phenomena---often characterized by shared covariance terms between $x_1$ and $(x_2 - x_1)$---is largely a consequence of the choice of coordinate system. While analyzing change ($d$) against baseline ($x_1$) conflates arithmetic coupling with measurement error effects, these issues are conceptually independent. For instance, in an analysis of absolute final states ($x_2$ vs. $x_1$), mathematical coupling is entirely absent, yet the RTM effect remains as a persistent bias driven by $\delta$. By maintaining this distinction, we emphasize that RTM is a physical consequence of measurement uncertainty, whereas coupling is an interpretative byproduct of how change is defined.

\subsection*{The Crude Regression Slope as a Biased Estimate}

The slope of the linear regression of the measured change $d$ on the measured pre-test value $x_1$ provides the crude estimate, $\beta_c$. This estimate is defined by the ratio of the covariance between $d$ and $x_1$ to the variance of $x_1$ \citep{Wasserman_2004}
\begin{equation}
\beta_c = \frac{\mbox{cov}(d,x_1)}{\mathbb{V}(x_1)} .
\end{equation}
By expanding the covariance term using our structural definitions ($\mbox{cov}(x_2 - x_1, x_1)$) and following the derivation by \citet{Hayes_1988}, we obtain an explicit expression for the crude slope in terms of structural parameters
\begin{equation}\label{bc}
\beta_c =  \frac{\beta \gamma^2 -\delta^2}{\gamma^2 + \delta^2} =  \beta - \frac{\delta^2}{\mathbb{V}(x_1)} (1+\beta).
\end{equation}
A fundamental insight from this equation is that the population-level RTM effect on the crude slope is independent of the between-subject variation in the treatment effect ($\nu^2$). The bias becomes even more transparent when examining the discrepancy between the crude estimate and the structural estimand, $\beta_c - \beta$, 
\begin{equation}\label{bc-b}
\beta_c - \beta  =  - (1+ \beta) \frac{\delta^2}{\gamma^2 + \delta^2}.
\end{equation}
This result demonstrates that the bias in the crude slope is driven entirely by within-subject variation ($\delta^2$). In the context of our blood pressure simulation (Figure \ref{fig:1rev}B), where $\beta = 0$, the bias reduces to $R-1= -\delta^2/(\gamma^2 + \delta^2)$. Using our parameters ($\gamma=13.6, \delta=9.1$), this predicts a crude slope of $\beta_c \approx -0.31$, which deviates significantly from the true null of zero.  Notably, the magnitude of this bias is conditional on the true value of $\beta$. The bias is amplified for positive values of $\beta$ and diminishes as $\beta$ approaches $-1$. In the specific case where $\beta = -1$, the pre-test and post-test values ($x_1$ and $x_2$) become independent, and the bias vanishes. However, for most biological and clinical applications where $\beta > -1$, $\beta_c$ will consistently underestimate the true effect. 

In this section, we have established that the true biological effect ($\beta$) is frequently obscured by measurement error and biological noise. We have shown that the crude regression slope ($\beta_c$) is an inconsistent estimator of this effect. Having established this structural benchmark, we are now positioned to evaluate how traditional correction methods attempt---and sometimes fail---to recover $\beta$ from observed data.

 \section*{Methodological Corrections for  RTM}
 
 Having established that the crude regression slope $\beta_c$ is a biased estimator of the structural parameter $\beta$ in the presence of measurement error, we now evaluate methods designed to recover the true effect. These corrections generally fall into two categories: those that use the internal variance structure of the observed data (\citealt{Berry_1984}) and those that require external information regarding measurement repeatability (\citealt{Blomqvist_1977}).
 
\subsection*{The Berry et al.  method: Correction by Correlation}

Building on the bivariate normal framework, \citet{Berry_1984} proposed a method to counteract the RTM effect by introducing an adjusted change variable, $d_B$. This approach was later popularized in ecology by \citet{Kelly_2005}, who defined the adjusted change as
\begin{equation}\label{dB}
d_B  =   x_2 -  \hat{\mu}_2  - \hat{\rho}  \left (x_1 - \hat{\mu}_1 \right )
\end{equation}
where $\hat{\mu}_1, \hat{\mu}_2$ are sample means and $\hat{\rho}$ is the sample correlation between $x_1$ and $x_2$.

Despite its widespread adoption in various fields (\citealt{Chuang-Stein_1993, Gunderson_2023, Sudyka_2019}), a formal assessment of the Berry et al. method's performance within a structural model of change---specifically one accounting for both measurement error ($\delta$) and stochastic biological noise ($\nu$)---is currently lacking. To extend the understanding of this method, we provide a rigorous analysis of its properties by replacing sample estimates with their true population values. 
Substituting the population correlation $\rho$, which is defined by the structural parameters as
\begin{equation}\label{rho}
\rho = \frac{\mbox{cov}(x_1, x_2)}{\sqrt{\mathbb{V}(x_1) \mathbb{V}(x_2)}} = \frac{ (1+ \beta) \gamma^2 }{\sqrt{[(1+\beta)^2 \gamma^2 + \nu^2 +\delta^2] [\gamma^2 + \delta^2] }},
\end{equation}
we can calculate the resulting population slope of the regression of the adjusted change $d_B$ on the initial measured value $x_1$. This slope, $\beta_B = \mbox{cov}(d_B, x_1)/\mathbb{V}(x_1)$, is given by
\begin{equation}\label{bB}
\beta_B  = -\rho + \frac{(1+\beta) \gamma^2   }{\gamma^2 + \delta^2}.
\end{equation}

The bias in the Berry et al. method stems from its implicit assumption regarding the nature of biological change. Specifically, by substituting the expression for the population correlation (Equation (\ref{rho})) into the Berry formula (Equation (\ref{bB})) and solving for the parameters that satisfy $\beta_B = \beta$, it becomes evident that the terms with opposing signs in the Berry formula cancel out only if the system is in a `steady-state' of variance where $\mathbb{V}(x_1) = \mathbb{V}(x_2)$. In a structural context, this equilibrium occurs only under the restrictive conditions of $\beta = 0$ and $\nu^2 = 0$.

However, when biological noise exists ($\nu^2 > 0$), the Berry et al. adjustment treats this natural variation as if it were an RTM artifact, leading to an overcorrection. Furthermore, if the true structural effect is non-zero ($\beta \neq 0$), the method is inherently biased regardless of the value of $\nu^2$, as it cannot accommodate the change in variance that a true differential effect produces. Consequently, what is intended as a correction for statistical bias becomes a source of biological misinterpretation.

The relationship between the corrected and crude slopes can be expressed as
\begin{equation}\label{difBc}
\beta_B  =  \beta_c + ( 1 -\rho).
\end{equation}
Because the correlation coefficient $\rho$ is bounded by $[-1, 1]$, the term $(1 - \rho)$ is always non-negative. This implies that the Berry et al.  method always adjusts the crude estimate in a positive direction.This mathematical property leads to an important insight regarding the method's accuracy under different structural scenarios.  As we established earlier,  the crude slope $\beta_c$ is a biased estimate whose direction depends on the value of $\beta$:
\begin{itemize}
\item [-] If  $\beta > -1$,  the crude slope underestimates the true effect ($\beta_c < \beta$). In these cases, the positive shift of the Berry et al.  method ($1-\rho$) moves the estimate toward the true structural slope. For our blood pressure example ($\beta=0, \nu=10, \gamma=13.6, \delta=9.1$), the Berry et al.  method yields $\beta_B \approx 0.08$, which is closer to the true value of 0 than the crude slope ($\beta_c \approx -0.31$).
\item [-] If  $\beta < -1$,  the crude slope overestimates the true effect ($\beta_c > \beta$). Here, the Berry et al.  method's positive correction exacerbates the bias, pushing the estimate further away from the structural parameter. Under these specific conditions, the uncorrected crude slope $\beta_c$ remains a more robust estimator than the  corrected slope $\beta_B$.
\end{itemize}

\subsection*{The Blomqvist Method: Correction by Repeatability}

While the Berry et al.  method relies on observed correlations, an alternative approach proposed by \citet{Blomqvist_1977} uses information about the measurement repeatability to provide an unbiased estimate of the structural parameter $\beta$. This method is particularly relevant in clinical settings where the repeatability of a measurement (such as blood pressure or cholesterol) may be known from previous pilot studies or technical specifications.

The Blomqvist  method is designed to produce the true slope $\beta$ (\citealt{Blomqvist_1977}).  In fact, by rearranging the equation (\ref{bc})  for  the crude  slope  $\beta_c$,  we can express the true slope in terms of the crude slope
\begin{equation}\label{bv}
\beta = \beta_c \left ( 1 + \frac{\delta^2}{\gamma^2} \right ) +  \frac{\delta^2}{\gamma^2}   = \frac{\beta_c \mathbb{V}(x_1)  + \delta^2}{\mathbb{V}(x_1) - \delta^2}.
\end{equation}
As previously noted, this correction is of limited practical use because it requires knowledge of the measurement error variance $\delta^2$, which cannot be estimated from typical two-time point data.

An alternative way to understand the Blomqvist correction, more in the spirit of the Berry et al. method, is to consider an adjusted change $d_e$,  where the subscript `e' denotes that the method is designed to yield an exact or unbiased slope estimate. The adjusted change is 
\begin{equation}\label{de}
d_e =   x_2 -  \hat{\mu}_2+   B \left( x_1 - \hat{\mu}_1 \right ), 
\end{equation}
where $B$  is a parameter chosen to ensure that the regression of $d_e$ on $x_1$  yields the true slope $\beta$.  By setting 
$\beta = \mbox{cov}(d_e,x_1)/\mathbb{V}(x_1)$,  we can solve for the required value of $B$:
\begin{equation}\label{BBe}
B =  \frac{  \beta  \delta^2 - \gamma^2}{\gamma^2 +\delta^2} = (1+ \beta_c) \frac{\delta^2} {\gamma^2} - 1 = (1+ \beta_c) \frac{\delta^2} {\mathbb{V}(x_1) - \delta^2} - 1  .
\end{equation}
This shows that to apply the transformation that recovers the true slope, we need to know the crude slope $\beta_c$,  the variance of the initial values $\mathbb{V}(x_1)$, and the measurement error variance $\delta^2$. 

Despite its mathematical superiority, the Blomqvist method is used less frequently than the Berry et al. method in organismal ecology. This is due to the requirement of an external estimate for $\delta^2$ (or the repeatability coefficient $R = \gamma^2/\mathbb{V}(x_1)$).   As we noted earlier,  $\delta^2$ cannot be determined from two-time-point data alone.  Therefore, the choice between these methods involves a trade-off between practical convenience and structural accuracy. The Berry et al.  method is data-sufficient  but prone to over- or under-correction depending on the biological noise, whereas the Blomqvist method is structurally robust but requires prior knowledge of the measurement's repeatability. This underscores the fundamental necessity of accounting for measurement repeatability in longitudinal studies; it is the only way to reliably bridge the gap between observed change and a causal understanding of the structural model. In designs restricted to a single baseline and follow-up, this requires an external estimate of error variance or a third measurement to partition within-subject noise from true biological change (\citealt{Chiolero_2013}).

\section*{Analysis of the Population Regression Slopes}  

Before we evaluate the regression slopes graphically, it's instructive to analyze the population values in the limiting cases of zero ($\delta^2 =0$) and  infinite  ($\delta^2  \to \infty$) measurement error variance.  This theoretical analysis provides a data independent assessment of the methods' intrinsic properties.

For $\delta^2=0$,  equation (\ref{bc}) yields $\beta_c=\beta$, as expected since in this case there is no regression to the mean.  However,   setting $\delta^2=0$ in equation (\ref{bB}) yields 
\begin{equation}
\beta_B =  \beta  + 1 -  \frac{\mbox{sgn}(1+\beta)}{\sqrt{1 + \nu^2/[(1+\beta)^2\gamma^2]}}.
\end{equation}
This result  shows that Berry et  al.  method  gives the correct slope (i.e.,  $\beta_B =  \beta$) only for $\nu^2 =0$ and $\beta > -1$. In particular, for $\nu^2 =0$ and $\beta < -1$ we have  $\beta_B =  2 +\beta$.  This is a critical finding, as it demonstrates that the Berry et al. method introduces a bias when no correction is needed, producing completely spurious results.

For the opposite limit, as $\delta^2 \to \infty $,   the measurement noise overwhelms the true biological signal.  In this case,  equation (\ref{bc}) yields $\beta_c \to - 1$. This is a sensible result,  as the measured data points $x_1$ and $x_2$  become effectively independent in this limit.  In contrast,  equation (\ref{bB}) yields $\beta_B \to 0$.  This implies that the Berry et al. correction misinterprets the noise-dominated data as representing an underlying relationship with no differential treatment effect ($\beta = 0$).  Of course,  the true underlying relationship between $X_1$ and $X_2$  is inaccessible from data in this limit.

To evaluate the continuous dependence of the slopes $\beta_c$ and $\beta_B$ on the various parameters of our framework, we conducted a simulation study using the same empirical values established for the blood pressure example in Figure \ref{fig:1rev}. Following \citet{Gardner_1973}, we set the baseline parameters at $\mu = 141$ mmHg and $\gamma = 13.6$ mmHg, with a within-subject (error) standard deviation of $\delta = 9.1$ mmHg.  Since the structural parameters $\alpha$ and $\beta$ must be set for the simulation, we fixed $\alpha = -20$ mmHg (following \citet{Hayes_1988}) and varied $\beta$ to observe its effect on the resulting estimates. The between-subject treatment variation was set to $\nu = 10$ mmHg. Notably, the derived slopes do not depend on either the population mean, $\mu$, or the additive treatment effect, $\alpha$, demonstrating that the systematic biases identified in our framework are general properties of the variance structure and the structural slope $\beta$.

%----------------------------------------------------------
\begin{figure}[h!]
\centering
  \includegraphics[width=\textwidth]{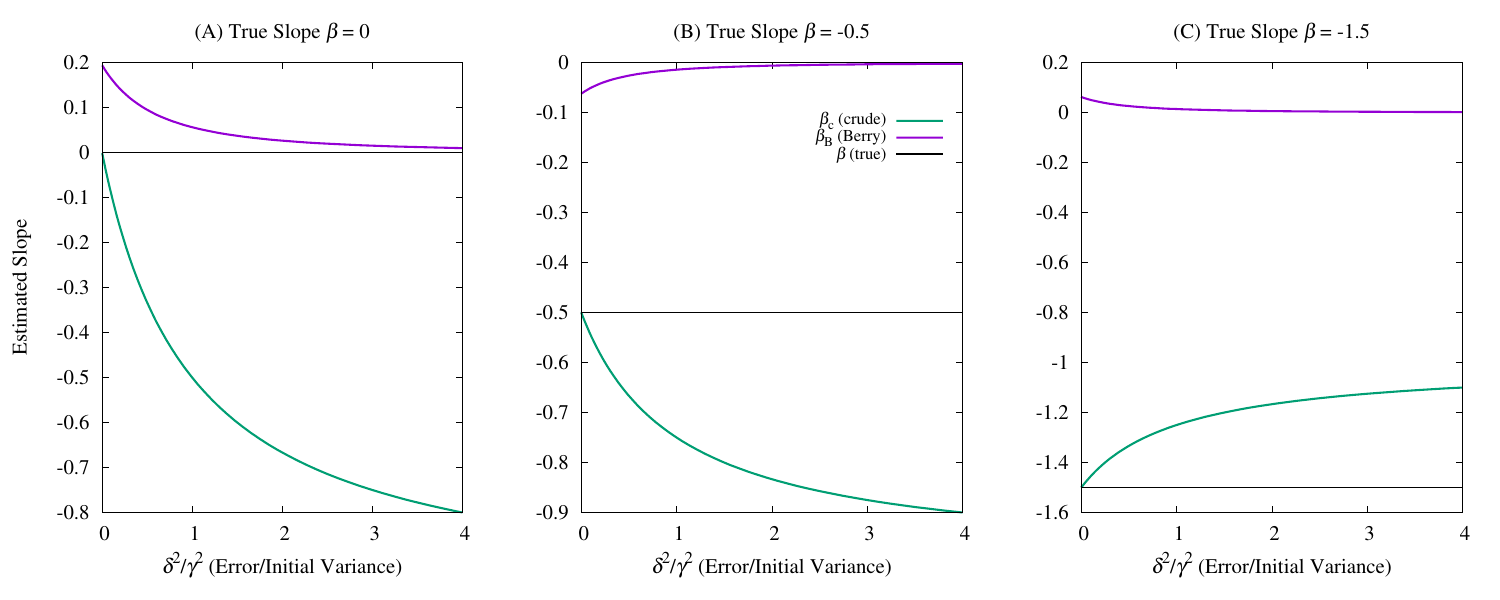}
\caption{Comparison of estimated slopes as a function of measurement noise. The three panels illustrate the behavior of the crude regression slope ($\beta_c$, green) and the Berry et al. adjustment ($\beta_B$, purple) compared to the true structural slope ($\beta$, horizontal black line). The x-axis represents the ratio of measurement error variance to initial true variance ($\delta^2/\gamma^2$), where a value of 0 indicates perfect repeatability and higher values indicate increasing measurement noise. In panel (A), which represents the structural null ($\beta = 0$), the crude slope is biased toward $-1$ as noise increases while the Berry et al.  estimate remains near the truth. In panel (B), where a moderate effect exists ($\beta = -0.5$), both methods diverge from the truth in opposite directions. In panel (C), representing a strong effect ($\beta = -1.5$), the crude slope approaches $-1$ while the Berry et al.  method overcorrects toward zero. All simulation parameters are based on the blood pressure model system ($\mu = 141$ mmHg, $\gamma = 13.6$ mmHg, $\nu = 10$ mmHg).}
\label{fig:2rev}
\end{figure}
%----------------------------------------------------------

The observed and adjusted slopes vary as a function of the ratio between measurement error and between-subject variance ($\delta^2/\gamma^2$; Fig. \ref{fig:2rev}). This ratio is directly related to repeatability ($R = 1/(1 + \delta^2/\gamma^2)$), a measure of consistency. For instance, the empirical ratio for systolic blood pressure data is approximately $\delta^2/\gamma^2 \approx 0.45$, which corresponds to $R \approx 0.69$. Given that measurement error variance $\delta^2$ is the mechanical driver of the RTM effect but is rarely measured in two-time point studies of change (\citealt{Chiolero_2013}), we treat it as the primary independent variable in our analysis.

We have not included the Blomqvist method in the preceding analysis because, in theory, it is designed to yield the true slope regardless of the parameter values. However, as we will demonstrate, this method's efficiency is severely constrained by limited sample size. We will show that this can cause the Blomqvist method to produce estimates of the true slope that are worse than the crude estimate, a counter-intuitive finding that highlights the method's practical limitations.

\section*{Sample Size Effects on Regression Slopes}

Equations (\ref{bc}),  (\ref{bB}), and (\ref{bv}) provide the population values for the crude,  Berry et al.,  and  Blomqvist   regression slopes.  While their simplicity allows for a complete assessment of the biases as a function of the model's parameters, a practical study relies on a sample of individuals. Consequently, the observed regression slopes calculated from a sample will inevitably differ from these population values due to sampling variation. In this section, we investigate the impact of this sampling variation and quantify its effect on the accuracy of the estimated slopes.

Using the parameters for systolic blood pressure  (\citealt{Gardner_1973,Hayes_1988}), 
we generate a sample of size $N$  by first drawing the initial (or baseline) true value $X_1$ from a normal distribution,  $X_1 \sim N(\mu,\gamma^2)$.  The final (or post-treatment) true value,  $X_2$,  is then generated using Equation   (\ref{X2}),  where $X_2=D+X_1$,   with noise $ \zeta \sim N(0,\nu^2)$.   Once the true values  $X_1$ and $X_2$ are known,  we  generate the observable values $x_1$ and $x_2$ using Equations (\ref{xx1}) and (\ref{xx2}) with  measurement error $\epsilon_i \sim N(0,\delta^2)$ for $i=1,2$. This procedure is repeated $N$ times to create a sample, from which we can directly calculate the regression slopes. 
We also define the slope of the regression of the true change,   $D=X_2-X_1$,   on the true initial value,  $X_1$,  as $\beta_t$.  While $\beta_t = \beta$ for an infinitely large sample,  it will generally differ for a sample of finite size $ N$. 

%----------------------------------------------------------
\begin{figure}[h!]
\centering
    \includegraphics[width=\textwidth]{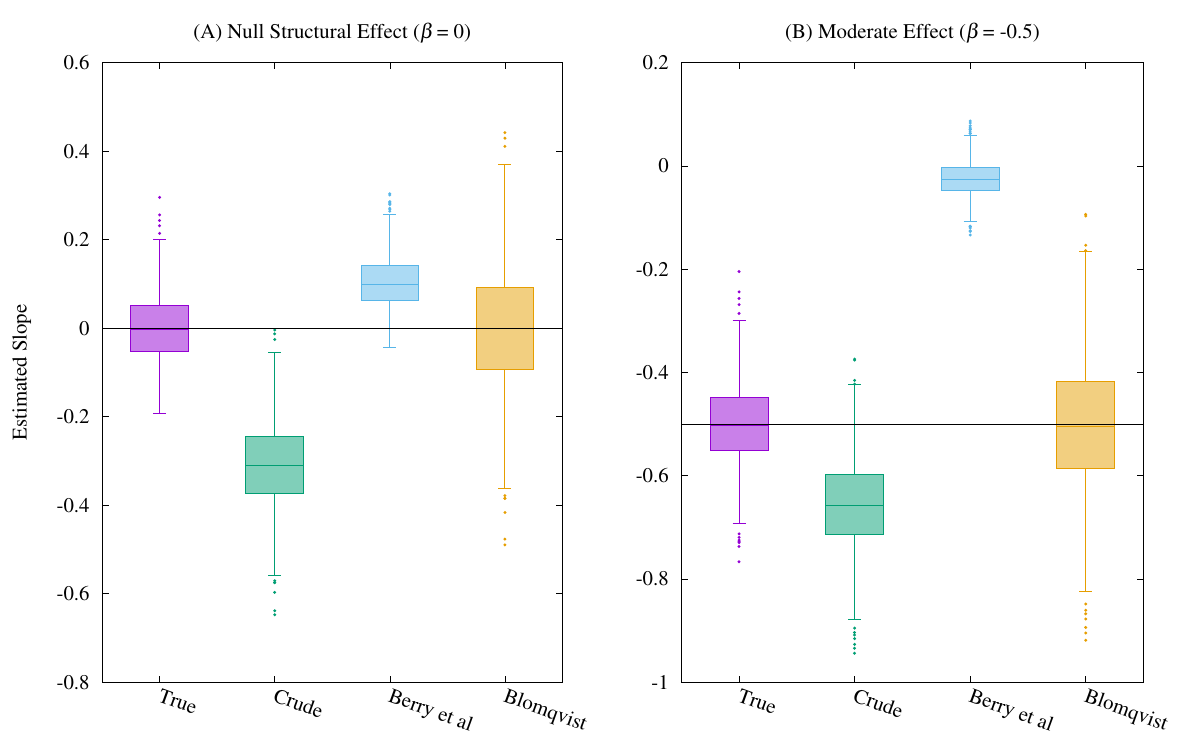}
\caption{Sampling distributions of regression slope estimators. Box plots represent the distribution of $10^3$ simulated slopes for a null structural effect ($\beta = 0$, panel A) and a moderate negative biological effect ($\beta = -0.5$, panel B).  In each plot, the central line denotes the median, the box bounds the interquartile range (IQR), and the whiskers encompass the data within 1.5 times the IQR. Individual points represent outliers beyond this range.  The horizontal black line indicates the true biological effect ($\beta$). Results demonstrate that even under a null effect, the crude ($\beta_c$) and Berry et al. ($\beta_B$) estimators remain biased by measurement error ($\delta$), while the Blomqvist ($\beta_e$) and true ($\beta_t$) slopes successfully recover the latent structural parameter. All simulations use the blood pressure model system parameters ($N=100$ per simulation).
}
\label{fig:3rev}
\end{figure}
%----------------------------------------------------------

Figure \ref{fig:3rev} shows box plots representing  the distribution of the various regression slopes  obtained from $10^3$ independent samples of size $N=100$.   The results highlight the biases of the crude slope $\beta_c$ and the Berry et al.  slope $\beta_B$,  as predicted by our population analysis.  The unexpected and critical finding is the large dispersion of the unbiased Blomqvist estimate $\beta_e$.  As a result,  for a given sample, this method can produce estimates that are farther from the true slope $\beta$ than the crude estimate.   We find the sampling variance of the Blomqvist estimate to be approximately $\mathbb{V}(\beta_e) \approx 0.020$ for $\beta=0$ and  $\mathbb{V}(\beta_e) \approx 0.016$ for $\beta\approx -0.5$. These values  are approximately twice  the variance of the crude slope.

This result underscores a significant limitation of the Blomqvist method, which already requires the often-unavailable a priori knowledge of the measurement error variance. Mathematically, the variance of the estimator, $\mathbb{V}(\beta_e)$, scales inversely with sample size ($1/N$). This property follows from the estimator's consistency, as formally demonstrated by \citet{Blomqvist_1977}, ensuring that the estimate converges to the true structural slope as $N$ increases. Nevertheless, the high sampling variance at the sample sizes typical of many ecological studies remains a formidable practical hurdle for robust inference. Indeed, while the Blomqvist estimator is consistent in the limit, our analysis of the Mean Squared Error (see Supplementary Material) reveals that its high sampling variance often outweighs its benefit as an unbiased estimator at small sample sizes.  
Specifically, at modest sample sizes ($N < 50$), the Blomqvist estimator becomes numerically unstable, producing extreme outliers when sampling fluctuations cause the observed variance to approach the known measurement error. Consequently, for the typical range of sample sizes in ecological research, biased but more precise alternatives often provide a more reliable and efficient estimate of the true biological effect.

\section*{Testing for a Differential Treatment Effect}

Our analysis demonstrates that regression to the mean (RTM) complicates the estimation of the structural relationship between change and initial values. Because measurement errors are ubiquitous and often difficult to quantify independently (\citealt{Castaneda_2012}), researchers face a persistent challenge in parameter estimation.  A common and perhaps more constrained objective is to assess the extent to which the observed data are compatible with a null structural slope ($\beta = 0$), representing a scenario where systematic individual differences in treatment response are negligible.  

The central challenge in hypothesis testing for $\beta=0$ is that the crude slope ($\beta_c$)
 is a biased estimate of the true slope ($\beta$).  As shown in equation (\ref{bc}),  under the null hypothesis that $\beta=0$, 
 the crude slope has a negative bias
 \begin{equation}\label{null_bc}
 \beta_c = -\frac{\delta^2}{\gamma^2 + \delta^2} =- \frac{\delta^2}{\mathbb{V}(x_1)} .
 \end{equation}
Consequently, even in the case of a null structural slope ($\beta = 0$), the regression of change on initial values will still yield a negative observed slope. A researcher who is unaware of the RTM effect and simply tests whether the observed slope $\beta_c$ is different from zero may falsely reject the null hypothesis, potentially attributing the result to a systematic differential treatment effect that is not supported by the underlying structural model.
 
 Therefore, the correct null hypothesis is that the observed crude slope is statistically equal to $-\delta^2/\mathbb{V}(x_1)$ or, equivalently,  to $R-1$ if we use the repeatability $R$.  However, as we have noted, this approach may  not  be practical because it requires knowing the measurement error variance, $\delta^2$,  or the repeatability $R$,  which are rarely available in two-time point studies. Nevertheless, if a qualitative assessment of the value of $R$ can be made,   this method can be valuable, as we will demonstrate next. 

 The Berry et al. method, despite its intuitive appeal as a correction for the RTM effect, presents significant drawbacks when used for hypothesis testing. As shown in our population analysis, under the null hypothesis that $\beta=0$, the corrected slope is
 \begin{equation}\label{null_B}
 \beta_B = \frac{\delta^2}{\gamma^2 + \delta^2}  \left [  \frac{1}{\sqrt{1 + \nu^2/(\gamma^2 + \delta^2)}}-1 \right ]= \frac{\delta^2}{\mathbb{V}(x_1)}  \left [  \frac{1}{\sqrt{1 + \nu^2/\mathbb{V}(x_1)}}-1 \right ] .
 \end{equation}
This expression is always negative for $\nu^2 > 0$.  For instance,  when $\nu^2 \ll \mathbb{V}(x_1)$,  the slope can be approximated as $\beta_B \approx - \delta^2 \nu^2/[2 (\mathbb{V}(x_1))^2]$. This shows that the corrected slope is systematically influenced by the stochastic between-subject variation in the treatment effect ($\nu^2$).   Consequently, a researcher utilizing this method might observe a non-zero slope even when the structural differential treatment effect is null, increasing the probability of a false rejection of the null hypothesis.

Furthermore, the behavior of the Berry et al. estimator in the presence of substantial measurement error poses a significant risk of Type II error.   As the measurement error variance ($\delta^2$) increases relative to the between-subject variance ($\gamma^2$), our analysis demonstrates that the adjusted slope $\beta_B$ mathematically converges toward zero, regardless of the true biological effect $\beta$ (see Figure \ref{fig:2rev}).   In such scenarios, a researcher may obtain a slope near zero and fail to reject the null hypothesis, even when the systematic treatment effect is substantial. This convergence makes the Berry et al. method unreliable for estimating structural biological effects, as the resulting estimate is driven more by the magnitude of measurement noise than by the underlying biological process.

%----------------------------------------------------------
\begin{figure}[h!]
\centering
    \includegraphics[width=\textwidth]{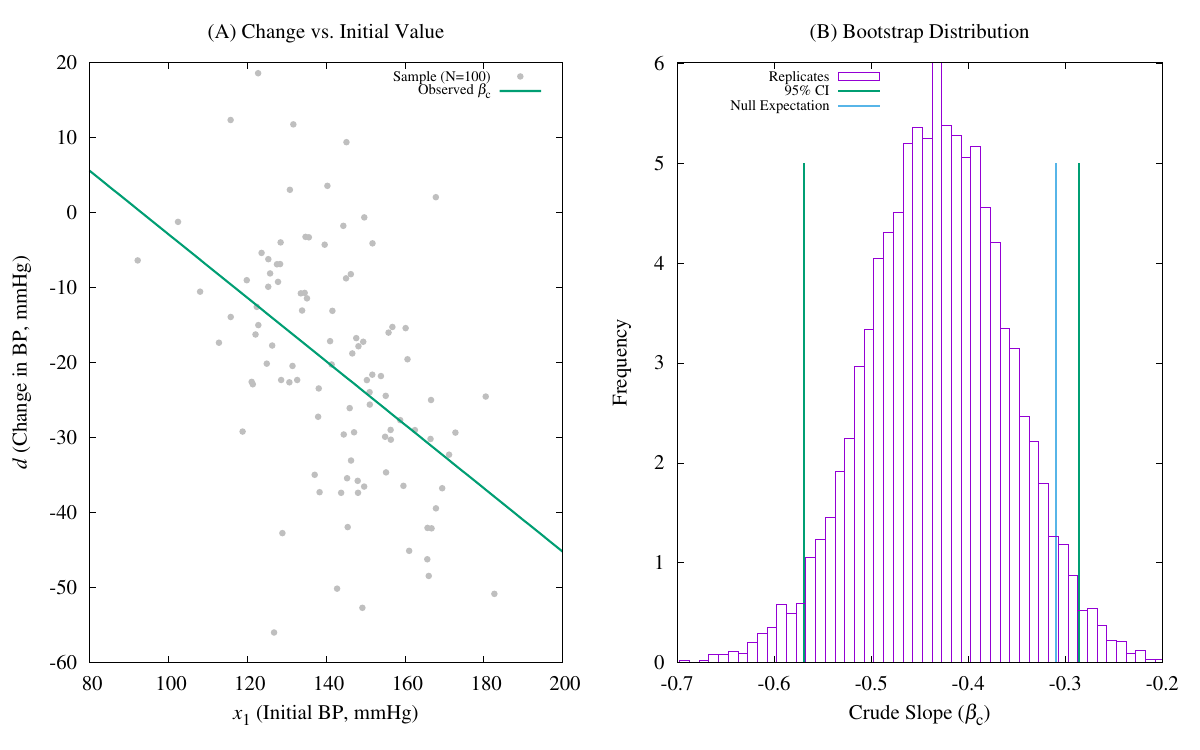}
\caption{Procedure to test for a null structural effect ($\beta=0$) using the crude slope. Using a simulated sample of systolic blood pressure ($N=100$), panel (A) shows the scatter plot of change ($d$) against initial value ($x_1$). The resulting empirical crude slope (green line) is $\beta_c = -0.423$. Panel (B) shows the histogram of $10^4$ crude slopes generated by bootstrapping the empirical sample. The vertical green lines indicate the limits of the $95\%$ bootstrap confidence interval ($[-0.569, -0.286]$). The vertical light-blue line indicates the expected null slope ($\beta_c = -0.31$), which is the association expected from measurement error alone when the true biological effect is null. Because the null value falls within the confidence interval, we fail to reject the hypothesis of a null structural effect. Simulation parameters: $\mu = 141$, $\gamma = 13.6$, $\delta = 9.1$, $\alpha = -20$, and $\nu = 10$.
}
\label{fig:4rev}
\end{figure}
%----------------------------------------------------------

To illustrate how we can test the null hypothesis $\beta=0$ using the crude slope, we use the data for systolic blood pressure to generate a single sample of size $N=100$. The data of change $d = x_2 - x_1$ against initial value $x_1$ is shown in Figure \ref{fig:4rev}. The empirical regression slope is $\beta_c =  -0.423$.  From our population analysis (Equation  (\ref{null_bc})), we know that the expected value for the crude slope under the null hypothesis is $\beta_c = - 0.31$. Of course, this value is unknown in a real experiment, since we only have access to the measured data $x_1$ and $x_2$.

To evaluate the $95\%$  confidence interval for our observed crude slope, we generate $10^4$
  crude regression samples by bootstrapping from our empirical sample (\citealt{Efron_1993}). The resulting bootstrap histogram is shown in Figure \ref{fig:4rev}.  The $95\%$  confidence interval is $[-0.569,-0.286]$. This means that the null hypothesis $\beta=0$  cannot be rejected if the expected value of the crude slope, which is $R-1$, falls within this interval. The repeatability for the systolic blood pressure data is $R\approx 0.69$, which corresponds to a null value of $R-1=-0.31$.  Since $-0.31$  falls within the calculated confidence interval, the null hypothesis cannot be rejected.

However, in most cases, the repeatability $R$ is unknown. It is therefore left to the researcher to subjectively evaluate if the expected repeatability of the experiment is within the required confidence interval.  We recall that the efficacy of the bootstrap is strongly dependent on the quality of the empirical sample. In that sense, it is preferable to test the null hypothesis using the crude slope, rather than the Blomqvist slope, which has a much wider variation.

\section*{Biological Case Studies}

To demonstrate the practical implications of our structural framework, we re-examine two high-profile case studies in evolutionary biology. In these fields, researchers are often interested in individual plasticity---how much an organism can change its traits in response to the environment---and how that change depends on its starting state.
 
\subsection*{Heat Tolerance Plasticity in Lizards}

In thermal biology,  the  tolerance-plasticity trade-off hypothesis  suggests that individuals with high baseline heat tolerance have less physiological room to improve through acclimation.  Heat hardening is the process by which an organism increases its upper thermal limit after sub-lethal exposure to high temperature---a critical trait for surviving climate-driven heatwaves. \citet{Deery_2021} and \citet{Gunderson_2023} studied two lizard species (\textit{Anolis}) to determine if those already tolerant to heat showed lower plasticity ($\beta < 0$). This is a vital question: if the most tolerant individuals cannot harden further, they may be the most vulnerable to rising temperatures.

%----------------------------------------------------------
\begin{figure}[h!]
\centering
  \includegraphics[width=\textwidth]{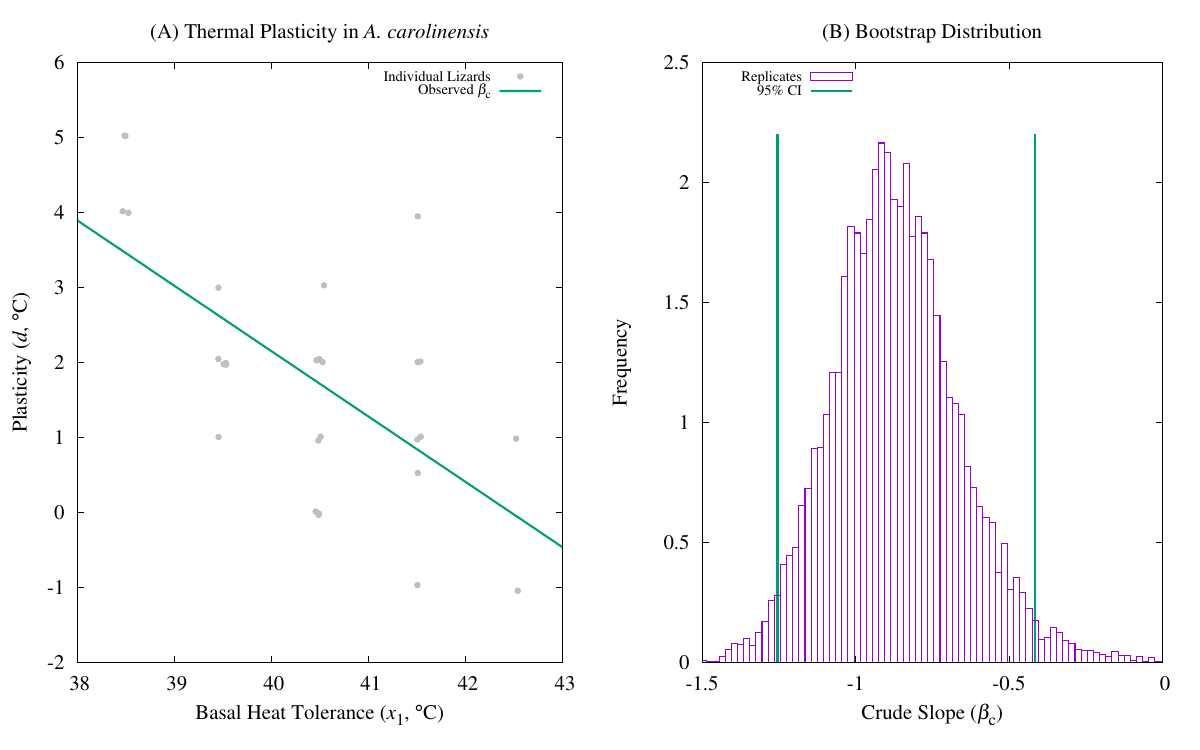}
\caption{Analysis of heat tolerance plasticity for the lizard \textit{Anolis carolinensis}. (A) Scatter plot of heat tolerance plasticity ($d$, the change in tolerance) against basal heat tolerance ($x_1$) for $N=30$ individuals. To reveal overlapping data points resulting from measurement rounding, a small amount of random jitter has been added to the point positions. The empirical regression (green line) shows a steep negative slope of $\beta_c = -0.872$. (B) Histogram of $10^4$ crude slopes obtained via bootstrapping. The vertical green lines indicate the $95\%$ confidence interval boundaries ($[-1.255, -0.415]$). This analysis replicates the standard approach in the literature, which often interprets such negative slopes as evidence of a biological trade-off or `compensation, 'though our structural model suggests this result is heavily influenced by regression to the mean.
}
\label{fig:5rev}
\end{figure}
%----------------------------------------------------------

Our re-analysis of thermal tolerance plasticity in \textit{Anolis carolinensis} ($N=30$) yields an empirical crude slope of $\beta_c = -0.872$, with a bootstrap $95\%$ confidence interval of $[-1.255, -0.415]$ (Fig. \ref{fig:5rev}). This strong negative association between observed change and basal tolerance initially appears to suggest a significant biological trade-off.

Using our structural model, we find that previous conclusions based on permutation tests (\citealt{Deery_2021, Gunderson_2023, Santos_2025}) are a clear example of a statistical pitfall. Permutation tests are typically used to test the hypothesis that $x_1$ and $x_2$ are uncorrelated, which in our framework corresponds to $\beta = -1$. This is a problematic null hypothesis for biological plasticity because it actually presupposes an extremely strong treatment effect---one where the final state is entirely independent of the initial state.

Our re-analysis shows that a null structural effect ($\beta = 0$) cannot be rejected if the repeatability ($R$)---the ratio of between-subject variance to total variance---falls in the range $R \in (0, 0.585]$. However, if these lizards have repeatable heat responses on the same order as that for human systolic blood pressure ($R \approx 0.69$), the null hypothesis should be rejected (Fig. \ref{fig:5rev}B). This demonstrates that accurate estimation of the structural parameter $\beta$ is not merely a statistical exercise; it determines whether we conclude a species is at its physiological limit or possesses a robust capacity for adaptation.

\subsection*{Telomeres as Biomarkers of Individual Quality}

Telomeres are protective protein-DNA caps at the ends of chromosomes that shorten as an animal ages. The rate of this shortening, known as telomere attrition, is widely used in evolutionary ecology as a biomarker of quality---a metric to predict an individual’s remaining lifespan and future reproductive success. A central question in aging research is whether individuals with longer initial telomeres experience faster attrition (a high biological aging rate).

\citet{Sudyka_2019} investigated this relationship in the Eurasian blue tit (\textit{Cyanistes caeruleus}) focusing on how initial telomere length ($\mbox{TL}_1$) influences the change toward the final measurement ($\mbox{TL}_{\mbox{\tiny last}}$). For consistency with our framework, we define the crude attrition rate as $-d = \mbox{TL}_1 - \mbox{TL}_{\mbox{\tiny last}}$.  In this context, the structural parameter $\beta$ represents the biological rate of aging relative to the initial state. If $\beta$ is estimated inaccurately due to RTM, researchers might mistakenly conclude that high quality individuals (those with long telomeres) are aging at a different rate than they truly are.

%----------------------------------------------------------
\begin{figure}[h!]
\centering
   \includegraphics[width=\textwidth]{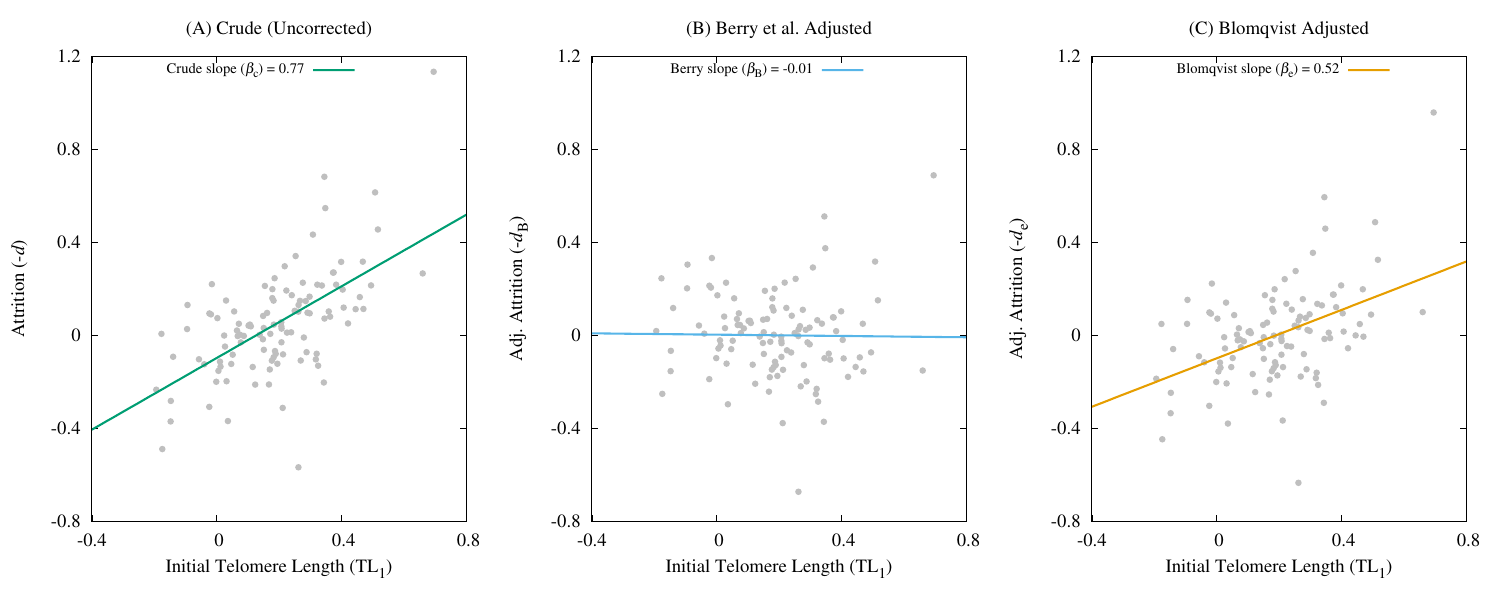}
\caption{Telomere attrition rates adjusted for the Regression to the Mean (RTM) effect. Data represent telomere length (TL) dynamics in the blue tit (\textit{Cyanistes caeruleus}), where attrition is expressed as the negative change ($-d = \mbox{TL}_1 - \mbox{TL}_{\mbox{\tiny last}}$) against initial telomere length ($\mbox{TL}_1$). (A) The uncorrected crude analysis suggest that longer telomeres shorten faster ($\beta_c = 0.770$). (B) The Berry et al. adjustment removes this correlation ($\beta_B = -0.014$), suggesting attrition is independent of initial length. (C) The Blomqvist adjustment results in a moderate correlation ($\beta_e = 0.520$). The stark disagreement between these  corrected  slopes highlights that without a structural model, different statistical adjustments lead to opposing biological conclusions about whether telomere shortening is dependent on initial length.
}
\label{fig:6rev}
\end{figure}
%----------------------------------------------------------

Figure \ref{fig:6rev} shows scatter plots of the crude attrition rate ($-d$), the Berry et al.  adjusted rate  ($-d_B$), and the Blomqvist adjusted rate ($-d_e$) in the whole dataset ($N=111$ birds; see the supplementary material in \citet{Sudyka_2019}). While  $-d$ and  $-d_B$   can be evaluated with the data available, the Blomqvist adjustment $-d_e$  requires knowledge of the  repeatability in telomere length.
Two estimates are provided in Table 1 in \citet{Kärkkäinen_2022}:  $R=0.479$ and $R= 0.398$.  The observed variance in initial measured telomere length  is $\mathbb{V}(x_1) =0.0309$.   Assuming $R=0.479$, we can estimate the component variances as $\gamma^2 = 0.0148$ and the measurement  error variance $\delta^2 = 0.0162$.  These results allow for the use of equations (\ref{de}) and (\ref{BBe}) to obtain $-d_e$.

%----------------------------------------------------------
\begin{figure}[h!]
\centering
  \includegraphics[width=\textwidth]{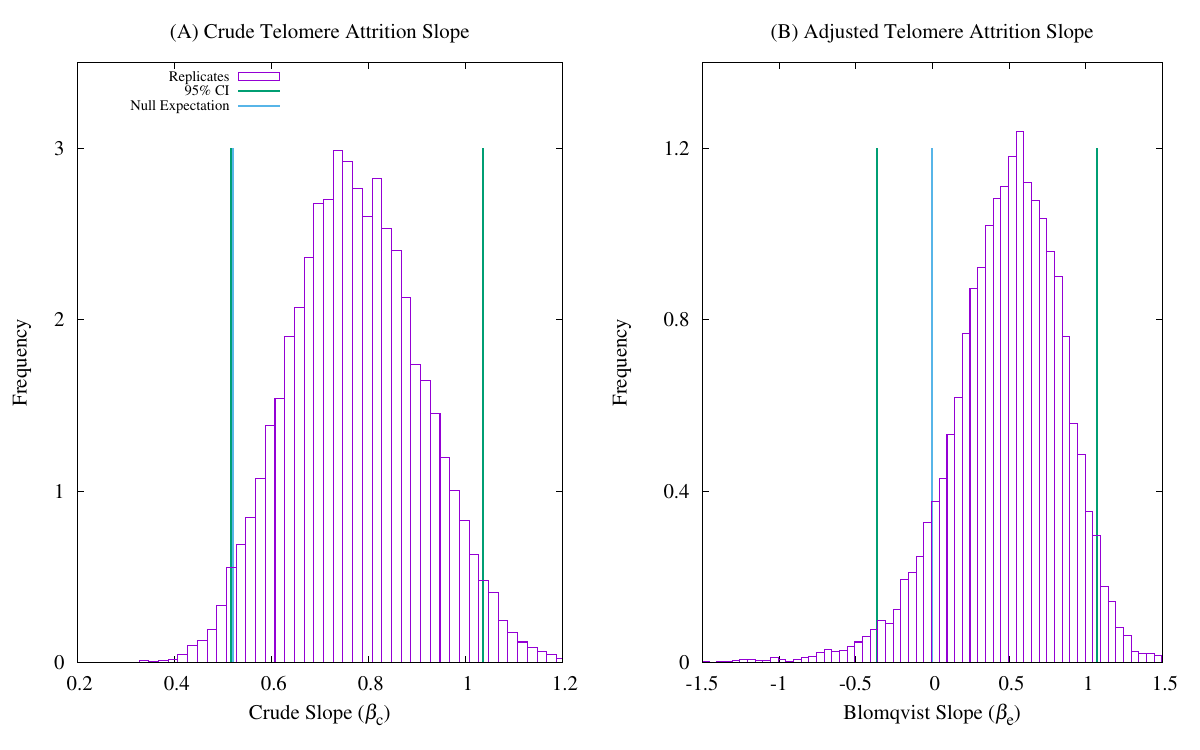}
\caption{Bootstrap distributions for telomere attrition slope estimates. The histograms represent $10^4$ bootstrap replicates of the regression slopes for telomere length (TL) change against initial TL. (A) Distribution of the crude slope ($\beta_c$). The vertical green lines indicate the $95\%$ confidence interval ($[0.517, 1.037]$). The vertical light-blue line indicates the null hypothesis expectation ($\beta_c = 0.521$) derived from measurement error. Because the null expectation falls within the confidence interval, the apparent relationship between initial length and attrition is statistically indistinguishable from regression to the mean. (B) Distribution of the Blomqvist adjusted slope ($\beta_e$). The $95\%$ confidence interval ($[-0.322, 1.075]$) contains the structural null value of zero (vertical light-blue line), leading to the same biological conclusion: there is no statistical evidence for a non-random relationship between initial telomere length and the rate of attrition in this sample.  
 }
\label{fig:7rev}
\end{figure}
%----------------------------------------------------------

We recall that knowledge of the measurement  error variance ($\delta^2$)  allows us to test the no-treatment-effect null hypothesis $\beta=0$ for the crude change by comparing the empirical result $\beta_c = 0.770$ with the null hypothesis expectation $\beta_c = 0.521$ given by equation (\ref{null_bc}).   Since the Blomqvist method produces an unbiased slope estimation,   with a value of $\beta_e =0.520$  for the telomere length data,  the null hypothesis is $\beta_e=0$.   The Berry  et al.  method does not allow for a direct  hypothesis  test  because the  between subject treatment variance ($\nu^2$) is unknown,  a value necessary for using the equation (\ref{null_B}). 

Accordingly, Figure \ref{fig:7rev}  shows the bootstrap histograms for $\beta_c$ and $\beta_e$. The $95\%$ confidence interval for the crude estimate is $[0.517, 1.037]$  which  just barely contains  the null hypothesis expectation of $\beta_c = 0.521$.  In contrast, the 
$95\%$ confidence interval for the Blomqvist estimate is $[-0.322, 1.075]$, which  broadly includes  the null expectation of $\beta_e = 0$.  

The wide uncertainty produced by the Blomqvist method reinforces our conclusion that while this method is structurally robust and theoretically unbiased, it is of limited utility for identifying biological signals in studies with modest sample sizes. Furthermore, we emphasize that the Berry et al. estimate ($\beta_B \approx 0$) should not be compared directly to a null of zero. 
Instead,  it should be compared with the null hypothesis expectation given in equation (\ref{null_B}), which accounts for the method's inherent bias.
Without accounting for these structural foundations, researchers risk over-interpreting  corrected trends in biomarker research that may simply reflect statistical uncertainty.

\section*{Conclusion}

The ultimate cause of regression to the mean (RTM) is measurement error, which is directly related to a measure of precision known as repeatability (R). It is logically inconsistent to propose methods that correct for RTM without any information on the magnitude of this error, as is the case with the Berry et al.  (\citealt{Berry_1984,Kelly_2005}) method.

Our analysis demonstrates that framing the problem within a proper change framework eliminates the need for complex correction methods.   In the simpler, yet common, problem of testing the null hypothesis of no differential treatment ($\beta=0$),  knowledge of the crude regression slope and even a qualitative assessment of the repeatability can provide the only solid information to guide researchers.   A 2012 literature survey published by \citet{Wolak_2012} showed that the median repeatability of physiological and behavioral traits is below $0.5$ ($0.30$ and $0.48$, respectively), although there is a large dispersion. This indicates that the null hypothesis $\beta =0$ could not be rejected if the expected null value of the crude slope, which is around  $-0.70$ for physiological traits and $-0.52$  for behavioral traits, falls within the $95\%$ confidence interval of bootstrapped empirical values.

We therefore argue that characterizations of the magnitude and direction of a structural treatment effect are statistically unfounded if they are not supported by an analysis of the experiment's repeatability. The key to solid inference is to evaluate the observed data against the expected bias, a task that requires an independent estimate of repeatability. This step is logically prior to the quantification of effect sizes; without it, researchers risk calculating precise confidence intervals around results that are essentially structural artifacts. Instead of attempting to blindly  correct for the RTM effect, we argue that researchers should instead evaluate whether their observed results are statistically inconsistent with the biases that are inherent to their experimental design.

To illustrate this quantification of uncertainty when repeatability is unknown, we provide a profile likelihood analysis in the Supplementary Material.  By mapping the structural envelope of the slope $\beta$ across a range of empirical datasets---including synthetic blood pressure, lizard heat tolerance plasticity and telomere length---we show how the width of the resulting confidence intervals remains constrained by a plateau of non-identifiability. This analysis provides a frequentist framework for reporting the full range of effect sizes compatible with the data, acknowledging that increasing sample size alone cannot collapse the uncertainty inherent in the measurement process.

Looking forward, we propose that a critical next step for the field is a systematic re-examination of published studies that have relied on biased RTM corrections. Because methods like the Berry et al. adjustment can create the illusion of biological trade-offs or compensatory growth where none exist, many established findings may require re-evaluation using structural models and knowledge of repeatability. Furthermore, while we emphasize the robustness of the structural null test, the profile likelihood approach developed here can bridge this method with the 'New Statistics' focus on effect-size estimation. Such structural models offer a natural path toward Bayesian hierarchical models (\citealt{McElreath_2020,Hector_2021,Cumming_2024}), which can formally incorporate prior knowledge of measurement error to generate posterior distributions for the structural slope $\beta$.

\section*{Acknowledgments}

M.S. is funded by grant PID2024-162000NB-I00 from Ministerio de Ciencia, Innovación y Universidades (Spain).  J.F.F.  is partially supported by Conselho Nacional de Desenvolvimento Científico e Tecnológico (Brazil),  grant number 305620/2021-5.

\newpage

\begin{center}
\section*{Supplementary Material}
\end{center}

 \setcounter{figure}{0}
  \setcounter{equation}{0}
 \renewcommand{\thefigure}{S\arabic{figure}}
 \renewcommand{\theequation}{S.\arabic{equation}}

\section*{Evaluation of the Bias–Variance Trade-off via Mean Squared Error (MSE)}

In the main text, we focused primarily on the structural bias of the crude slope, the Berry et al. method, and the Blomqvist estimator. While "unbiasedness" is a desirable property in a statistical estimator, it does not account for the precision of the estimate. In practical research, a theoretically unbiased estimator may be less reliable than a biased one if its sampling variance is excessively high. To provide a more comprehensive evaluation of these methods, we here analyze their performance using the Mean Squared Error (MSE).

The MSE is a measure of the total error of an estimator. It is mathematically defined as the expected value of the squared difference between the estimated slope ($\hat{\beta}$) and the true structural parameter ($\beta$), 
\begin{equation}
MSE = \mathbb{E}[(\hat{\beta} - \beta)^2].
\end{equation}
A key property of the MSE is that it can be decomposed into the sum of the estimator's squared bias and its variance:
\begin{equation}
MSE = \mbox{Bias}^2 + \mbox{Variance} = (\mathbb{E}[\hat{\beta}] - \beta)^2 + \mathbb{V}(\hat{\beta})
\end{equation}
This decomposition illustrates the fundamental trade-off in statistical inference: attempts to eliminate bias (reducing the first term) often result in a significant increase in sampling variance (increasing the second term). For an ecologist, the ``best" method is typically the one that minimizes the total MSE, representing the highest probability of obtaining an estimate close to the truth from a single dataset.

\begin{figure}[t!]
\centering
    \includegraphics[width=\textwidth]{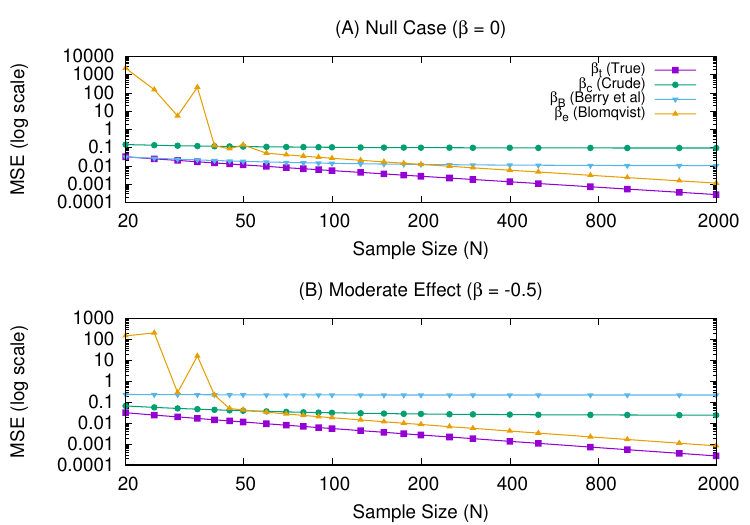}
\caption{ Evaluation of estimator performance via Mean Squared Error (MSE) across varying sample sizes ($N$). Results are shown for the true structural slope ($\beta_t$, purple), the crude slope ($\beta_c$, green), the Berry et al. method ($\beta_B$, sky blue), and the Blomqvist estimator ($\beta_e$, orange). The y-axis is presented on a logarithmic scale to accommodate the extreme variance observed at low sample sizes. (A) Null Case ($\beta = 0$). (B) Moderate Effect ($\beta = -0.5$). In both panels, the Blomqvist estimator ($\beta_e$) reveals a ``numerical instability zone” at small sample sizes ($N < 50$). In this region, sampling fluctuations in the observed variance can result in a denominator near zero, causing the MSE to spike by several orders of magnitude. For large $N$, the MSE of both unbiased estimates $\beta_t$ and $\beta_e$ vanish as $1/N$.
}
\label{fig:S1}
\end{figure}

To illustrate the trade-off between bias and variance,  Figure \ref{fig:S1}  shows the MSE for the three estimators using the parameters derived from the blood pressure study ($\nu=10$, $\gamma=13.6$, $\delta=9.1$). We simulated $10^6$ datasets for each sample size ($N$) across two scenarios: a null effect (Panel A, $\beta=0$) and a moderate effect (Panel B, $\beta=-0.5$). We also included the MSE of the structural slope ($\beta_t$)---calculated from the unobserved true values---as a ``gold standard" representing the theoretical lower bound of error achievable for a given $N$ and biological noise $\nu$.

As shown in Figure \ref{fig:S1}, the MSE of the two unbiased estimators ($\beta_t$ and $\beta_e$) vanishes at a rate of $1/N$ as sample size increases. Conversely, for the biased estimators ($\beta_c$ and $\beta_B$), the MSE approaches a non-zero constant defined by the square of their asymptotic bias.
Indeed,  since our population estimates of $\beta_c$ and $\beta_B$ are exact, we can write these asymptotic values as 
\begin{equation}
\mbox{MSE}(\beta_c) \to \left( \frac{(1+\beta)\gamma^2}{\gamma^2 + \delta^2} - (1+\beta) \right)^2
\end{equation}
and
\begin{equation}
\mbox{MSE}(\beta_B) \to \left( \frac{(1+\beta)\gamma^2}{\gamma^2 + \delta^2} - \frac{(1+\beta)\gamma^2}{\sqrt{(\gamma^2 + \delta^2)((1+\beta)^2\gamma^2 + \nu^2 + \delta^2)}} \right)^2 ,
\end{equation}
which match perfectly the plateaus observed in Figure \ref{fig:S1}.

In the null scenario shown in Figure \ref{fig:S1}A, the Berry et al. method ($\beta_B$) performs remarkably well, outperforming both the crude and Blomqvist estimators for $N < 200$. The reason is structural: the Berry method is specifically designed to shrink the slope toward zero (the null). When the true effect is indeed zero, this ``bias" aligns perfectly with the truth, effectively reducing sampling variance without introducing a centering error. In this specific case,  for small $N$,  $\beta_B$ is nearly as efficient as the gold standard $\beta_t$.

When a moderate effect is present, the scenario changes, as  shown in Figure \ref{fig:S1}B.  For very small samples ($N < 50$), the crude estimate ($\beta_c$) actually yields the lowest MSE among the three practical estimators. This is a classic example of the bias-variance trade-off: the crude method’s high bias is  ``cheaper" (in MSE terms) than the massive variance introduced by the Blomqvist correction or the mid-range bias of the Berry method at these low $N$ values.

A critical observation in our simulation is the erratic behavior of $\mbox{MSE}(\beta_e)$ for $N < 50$. Interestingly, increasing the number of Monte Carlo samples ($10^6$ to $10^8$) does not smooth these results; rather, it often produces even wilder fluctuations and higher MSE values.This occurs because the Blomqvist estimator is a ratio-based estimator where the denominator ($\mathbb{V}(x_1) - \delta^2$;  Equation (17) in the main text) can fluctuate arbitrarily close to zero. Statistically, this suggests that for small $N$, the distribution of $\beta_e$ has ``heavy tails" (similar to a Cauchy distribution), where the variance and mean are theoretically undefined. In a finite simulation, as you increase the number of samples, you increase the probability of hitting a ``near-singularity"---an iteration where the denominator is so small that the resulting $\beta_e$ is massive. Because MSE squares these outliers, a single ``near-zero" denominator can dominate the average of millions of simulations. This result strongly cautions against using Blomqvist-type corrections in small-sample ecological studies.

%----------------------------------------------------------
%----------------------------------------------------------

%\clearpage

\section*{Profile Likelihood for the Structural Model}

To complement the bootstrap approach and provide a more rigorous quantification of uncertainty, we implement a Profile Likelihood analysis (\citealt{Bolker_2008}). This method allows us to evaluate the plausibility of different values for the structural slope $\beta$ by considering the joint probability of the observed data under our structural assumptions.

Given $n$ pairs of observations $\mathbf{z}_i = [x_{1i}, x_{2i}]^T$, and assuming the data follow a bivariate normal distribution, the log-likelihood function is defined as
\begin{equation}
\ln L(\boldsymbol{\theta}) = -\frac{n}{2} \ln(2\pi) - \frac{n}{2} \ln |\mathbf{\Sigma}| - \frac{1}{2} \sum_{i=1}^n (\mathbf{z}_i - \boldsymbol{\mu})^T \mathbf{\Sigma}^{-1} (\mathbf{z}_i - \boldsymbol{\mu}) .
\end{equation}
Following the notation of our structural model, the parameter vector $\boldsymbol{\theta}$ consists of $\{\mu, \alpha, \beta, \gamma^2, \delta^2, \nu^2\}$. The mean vector $\boldsymbol{\mu}$ and the covariance matrix $\mathbf{\Sigma}$ are defined strictly by these structural parameters, 
\begin{equation}
\boldsymbol{\mu} = \begin{pmatrix} \mu \\ \mu(1+\beta) + \alpha \end{pmatrix}
\end{equation}
\begin{equation}
\mathbf{\Sigma} = \begin{pmatrix}
\gamma^2 + \delta^2 & (1+\beta)\gamma^2 \\
(1+\beta)\gamma^2 & (1+\beta)^2\gamma^2 + \nu^2 + \delta^2 
\end{pmatrix} .
\end{equation}
The profile likelihood for the slope $\beta$ is constructed by fixing $\beta$ at a specific value and maximizing the log-likelihood with respect to all other ``nuisance" parameters ($\mu, \alpha, \gamma^2, \delta^2, \nu^2$):
\begin{equation}
\ln L_{profile}(\beta) = \max_{\mu, \alpha, \gamma^2, \delta^2, \nu^2} \ln L(\beta, \mu, \alpha, \gamma^2, \delta^2, \nu^2) .
\end{equation}
This process is repeated over a range of $\beta$ values to produce the likelihood profile. Let $\tilde{\beta}$ denote the value of $\beta$ that maximizes this profile, representing the global Maximum Likelihood Estimate (MLE) for the structural slope. The 95\% confidence interval for $\beta$ is then defined by the set of values where the profile log-likelihood is within 1.92 units of this global maximum (\citealt{Bolker_2008}):
\begin{equation}
\ln L_{profile}(\beta) \geq \ln L_{profile}(\tilde{\beta}) - 1.92 .
\end{equation}
Note that this 1.92 threshold is derived from the $\chi^2$ distribution with one degree of freedom, corresponding to the critical value for a 95\% interval ($\chi^2_{1, 0.95}/2 \approx 3.84/2 = 1.92$).

As noted in the main text, with only two time points and no external information on repeatability ($R = \gamma^2 / (\gamma^2 + \delta^2)$), the structural model is under-identified. Specifically, $\delta^2$ and $\nu^2$ cannot be uniquely separated from the observed variances. In our implementation, we address this by exploring the parameter space across a biologically plausible range for the error variance $\delta^2$. The resulting profile likelihood provides a "structural envelope" of uncertainty, demonstrating that while multiple combinations of parameters can explain the data, they lead to a bounded range of plausible values for the structural slope $\beta$.

We note that in this linear structural framework, the means ($\mu$ and $\alpha$) are location parameters that do not influence the estimation of the structural slope ($\beta$) or the variance components ($\gamma^2, \delta^2, \nu^2$). Specifically, the maximum likelihood estimate of $\beta$ is invariant to translations of the data. Consequently, in our Profile Likelihood implementation, we fix $\mu$ and $\alpha$ at their respective sample maximum likelihood estimates ($\tilde{\mu} = \bar{x}_1$ and $\tilde{\alpha} = \bar{x}_2 - (1+\beta)\bar{x}_1$) and focus the computational scan on the structural parameters that define the system's covariance.

\begin{figure}[t!]
\centering
    \includegraphics[width=\textwidth]{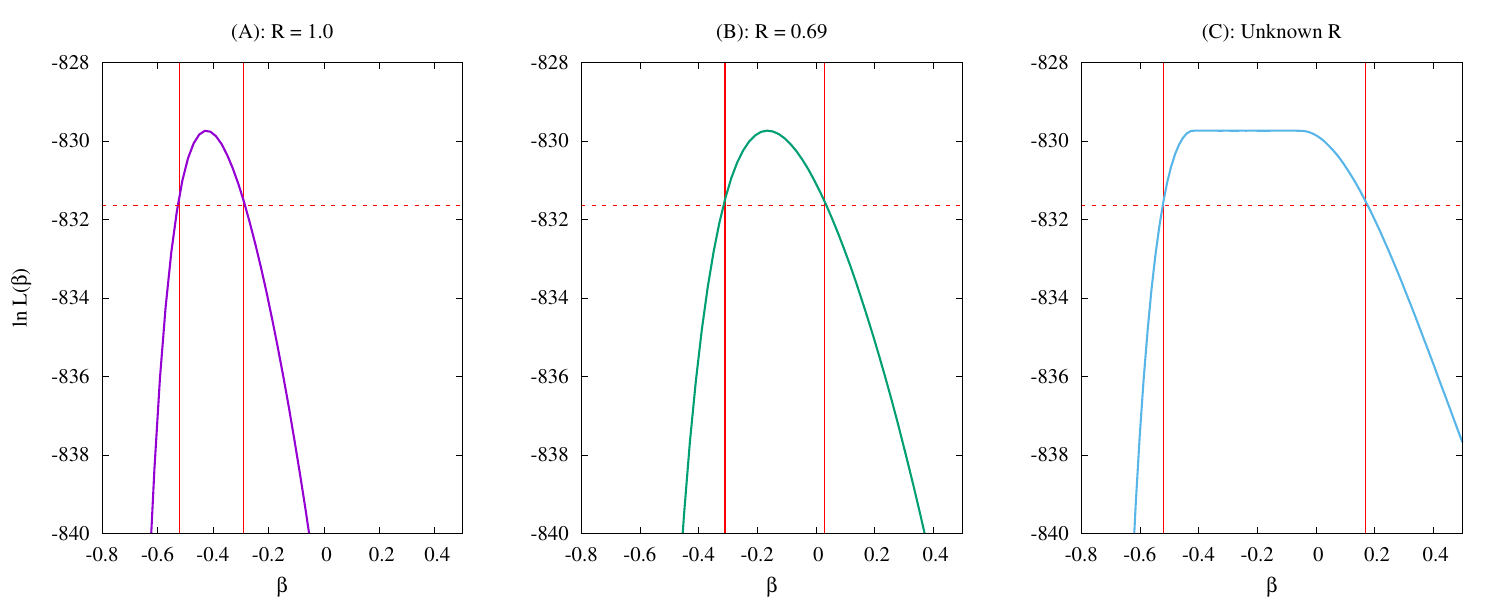}
\caption{ Profile Likelihood Analysis of the Structural Slope ($\beta$) for the systolic blood pressure data. The panels illustrate the log-likelihood profiles for the structural slope $\beta$ under different assumptions regarding the repeatability ($R$) of the baseline measurement. The horizontal red dashed lines indicate the 95\% confidence threshold, defined as $\ln L_{max} - 1.92$. Vertical solid lines indicate the boundaries of the 95\% confidence intervals.  (A) Neglecting Measurement Error ($R = 1.0$): Under the naive assumption that baseline measurements are error-free, the likelihood is maximized at $\tilde{\beta} \approx -0.42$. The resulting confidence interval is narrow but biased toward more negative values.  (B) Informed Structural Correction ($R = 0.69$): Using the known repeatability of the system, the profile shifts. The maximum likelihood estimate (MLE) is $\tilde{\beta} \approx -0.16$, and the confidence interval correctly encompasses zero, reflecting the true underlying null relationship.  (C) Profile Likelihood Over Unknown $R$: When $R$ is treated as an unknown nuisance parameter, the profile exhibits a ``likelihood ridge" or plateau. Because the model is under-identified, multiple combinations of $\beta$ and $R$ can produce the same maximum log-likelihood ($\ln L \approx -830$). The resulting ``structural envelope" provides a more honest quantification of total uncertainty, accounting for both sampling error and the lack of information regarding measurement precision.
}
\label{fig:S2}
\end{figure}

Figure  \ref{fig:S2} illustrates this analysis for our simulated dataset ($N=100, \beta_{true}=0, R_{true}=0.69$).  In panel A,  we assume measurement error is absent ($\delta^2 = 0$), so thar $R=1$.  Here, the likelihood is maximized at $\tilde{\beta} = -0.42$, and the 95\% confidence interval is $[-0.53, -0.28]$. This interval is narrow but entirely excludes the true value of zero, demonstrating the significant bias introduced by ignoring the regression to the mean effect.  In panel B, we use the informed repeatability ($R=0.69$) to correct the structural model. The likelihood peak shifts to $\tilde{\beta} = -0.16$, and the 95\% confidence interval is $[-0.32, 0.04]$. By correctly partitioning the variance, the estimate becomes unbiased and the confidence interval successfully encompasses the true value.  In panel C, we treat the repeatability $R$ as an unknown nuisance parameter. Because the structural model is under-identified with only two time points, the profile likelihood does not possess a single peak but rather a ``maximum likelihood ridge" or plateau. 
This plateau extends from $\beta \approx -0.42$ to $\beta \approx -0.08$.
Within this range, any value of the structural slope is equally supported by the data; for each $\beta$ in this interval, there exists a mathematically valid partition of the variance (a specific value of $R$) that allows the model to match the sample covariance matrix exactly, resulting in the same global maximum log-likelihood ($\ln L \approx -829.7$). The 95\% confidence interval  is $[-0.53, 0.17]$.  This ``structural envelope" illustrates that while the exact slope cannot be identified without prior knowledge of $R$, the data still place rigorous bounds on the range of plausible structural relationships.

%----------------------------------------------------------
%----------------------------------------------------------

\begin{figure}[t!]
\centering
    \includegraphics[width=\textwidth]{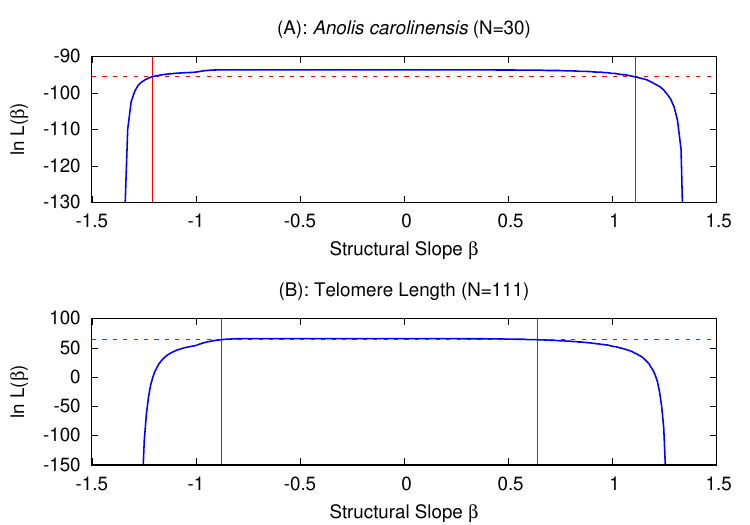}
\caption{ Profile likelihoods for the structural slope ($\beta$) under the assumption of unknown measurement repeatability ($R$). The structural relationship between the baseline trait and its subsequent change is shown for two empirical datasets: (A) heat tolerance plasticity in the lizard {\it Anolis carolinensis} ($N=30$) and (B) telomere length  as biomarkers of individual quality ($N=111$). In both panels, the solid blue line represents the profile log-likelihood maximized across all possible values of $R \in (0, 1)$. The characteristic ``plateau" at the peak of the likelihood indicates a zone of non-identifiability, where the data are equally consistent with a range of structural slopes by varying the assumed measurement error. The horizontal red dashed line indicates the 95\% confidence threshold ($\ln L_{max} - 1.92$), and the vertical red solid lines denote the boundaries of the resulting 95\% structural confidence intervals. Despite the larger sample size in panel B, the indeterminacy plateau remains a dominant feature, emphasizing that increasing sample size cannot resolve the bias of regression to the mean without independent information regarding measurement repeatability.
}
\label{fig:S3}
\end{figure}

For the {\it A. carolinensis}  dataset ($N=30$) (\citealt{Deery_2021}), the profile likelihood for the structural slope $\beta$ reveals a broad plateau of maximum likelihood extending from approximately $-0.88$ to $0.28$ (Figure \ref{fig:S3}A). Within this range, the model is under-identified; various combinations of the slope and the unknown repeatability $R$ fit the observed moments equally well. The resulting 95\% structural confidence interval, defined by a 1.92 log-likelihood drop from the plateau, is $[-1.21, 1.11]$. The inclusion of zero within the plateau indicates that the data are consistent with a null structural relationship once measurement error is accounted for.

Figure \ref{fig:S3}B shows that a similar pattern is observed in the analysis of the telomere length dataset ($N=111$) (\citealt{Sudyka_2019}). The maximum likelihood plateau for these data extends from $\beta \approx -0.77$ to $0.24$, and the 95\% structural confidence interval is $[-0.88, 0.64]$. Comparing the two panels of Figure \ref{fig:S3} reveals that  the width of the plateau  is not substantially reduced despite the larger sample size of panel B.  This persistence of the plateau demonstrates that the uncertainty in $\beta$ is not merely a consequence of low statistical power, but is a fundamental result of the mathematical under-identification of the repeatability $R$. Even with a larger sample size, the regression to the mean effect remains indistinguishable from a true biological relationship without an independent estimate of measurement reliability. In both empirical cases, the inclusion of $\beta = 0$ within the maximum likelihood plateau suggests that the observed correlations between baseline values and change can be parsimoniously explained by measurement error alone.

%----------------------------------------------------------
%----------------------------------------------------------

%\clearpage

\end{document}